# The Joint Diffusion of a Digital Platform and its Complementary Goods:

## The Effects of Product Ratings and Observational Learning

Meisam Hejazinia, Norris Bruce

Naveen Jindal School of Management, The University of Texas at Dallas,
meisam.hejazinia@utdallas.edu, norris.bruce@utdallas.edu,

## ABSTRACT

The authors study the interdependent diffusion of an open source software (OSS) platform and its software complements. They quantify the role of OSS governance, quality signals such as product ratings, observational learning, and user actions upon adoption. To do so they extend the Bass Diffusion Model and apply it to a unique data set of 6 years of daily downloads of the Firefox browser and 52 of its add-ons. The study then re-casts the resulting differential equations into non-linear, discrete-time, state space forms; and estimate them using an MCMC approach to the Extended Kalman Filter (EKF-MCMC). Unlike continuous-time filters, the EKF-MCMC approach avoids numerical integration, and so is more computational efficient, given the length of our time-series, high dimension of our state space and need to model heterogeneity. Results show, for example, that observational learning and add-on ratings increase the demand for Firefox add-ons; add-ons can increase the market potential of the Firefox platform; a slow add-on review process can diminish platform success; and OSS platforms (i.e. Chrome and Firefox) compete rather than complement each other.

INTRODUCTION

Software platforms are ubiquitous in markets for digital goods; famous examples are the *Linux* and *Windows* operating systems; and Firefox and Internet Explorer browsers. Here distinct participants contribute to innovation: platform owners who control access to platform services; and communities of third-party developers and end-users who create complementary software and generate online feedback. Platforms can be proprietary (Explorer) or "*open-source*" (Firefox); in the latter case the software is to be freely used, altered, and shared in agreement with specific licenses; and owners place fewer restrictions on who can participate in platform innovation (Eisenmann et al. 2009). As a novel approach to innovation (Lerner and Tirole 2002), open source software (OSS) stands as an important research topic (von Krogh and von Hippel 2006) in many disciplines. Yet empirical work has largely focused on factors that lead to OSS project success (e.g., time-to-launch) (Grewal et al. 2006; Stewart et al. 2006; Mallapragada et al. 2012), and less on the forces that account for OSS diffusion over time, among end-users. Moreover, diffusion studies in marketing have focused on proprietary platforms, where the central question is often how price and other marketing events (Shankar and Bayus 2003) moderate network externalities. What, however, accounts for the joint diffusion of an open platform and its complements, indeed software diffusion in the absence of pricing or other marketing incentives?

In this study, we draw upon the literature on OSS development, diffusion of innovation, and online user-generated content to address this our overall question. For example, owners of closed platforms typically employ strict internal review processes to screen complementary software for quality and user relevance (Evans et al. 2006); in contrast, in an open platform system (e.g., Linux, Apache) with limited central controls, end-user online feedback (Chintagunta et al. 2010) becomes one means through which potential adopters identify high-value (relevant) complements[1]. Similarly, successful open platform innovation will likely depend on the platform's ability to attract and retain third-party developers (Shah 2006); those who implement new complement features, fix bugs, and so on; which in turn could attract adopters to the platform (Gupta et al. 1999). Moreover, it is often these developers themselves who review software; and manage communication among peers (e.g., peer review process) and end-users. Thus, we draw on literature to consider how OSS self-governance may influence platform adoption over time. Finally, Raymond (1999) suggests that OSS projects should "release early, and release often" to produce higher quality software. Releasing a product early may attract developers, and gain early end-user feedback (Lakhani and Von Hippel 2003), which could then facilitate diffusion.

Thus, guided by the above issues and others to be outlined later, we consider the following research questions: To what extent do *open* complementary products influence platform adoption? What are effects of quality signals such as product ratings, observational learning, and user actions on the adoption of complementary software (plug-ins, add-ons, or complements)? How would the release strategy of an open platform and that of its complement jointly affect their diffusion? What is the effect of the complement's governing process on platform diffusion? Given our OSS platform context, we also control for the relative importance of licenses, developer rewards, and platform competition on diffusion.

To address these questions we extend the Bass diffusion model (hereafter BDM) to study the joint diffusion of an open digital platform and its complements, given BDM's parsimony, performance, and the intuitive interpretations of its parameters (Hauser et al. 2006). We apply the extended model to a canonical example, the diffusion of the Mozilla Firefox browser and its add-ons, an OSS (See Figure 1) project initiated in 1998 in response to Netscape's decision to publish its browser's source code. Our resulting response model is dynamic, nonlinear, and heterogeneous; thus for estimation, we re-cast BDM differential equations into non-linear, discrete-time, state-space forms; and then estimate them using an MCMC approach to the Extended Kalman Filter (EKF), a Sequential Monte Carlo method that works with normal-linear approximations. Our EKF-MCMC alogorithm is preferred to other Gaussian filters such as the Unscented Kalman Filter (UKF) and the Augmented Kalman Filter with continuous state space and discrete observations (AKF-CD) (e.g., Xie et al. 1997; Mahajan et al. 2000), given the lengths of our sample (T=1686 periods), the dimension of our state-space, and the importance of heterogeneity. Recall, the AKF-CD is continuous-time filter whose estimation requires a numerical integration at every period, and so it does not scale well to longer time series and/or high-dimensional problems, typically found in digital market studies. Moreover, the AKF-CD is a numerical procedure and so estimating heterogeneity is not trivial, as it is in iterative, sample based approaches (e.g., MCMC). Now unlike the EKF which approximates non-linear functions, the UKF instead approximates the posteriors of the state variables by sets of sample points (Ristic et al. 2004); and for highly nonlinear Gaussian systems the UKF produces typically better results; but our diffusion model *is quadratic* and so here EKF, performs just as well and is more computationally efficient.

---

[1] Both open and closed platforms help maintain software quality control via APIs and SDKs (Noori and Weis 2013). But open platforms often accept complements that are only relevant for only a small proportion of their installed bases.

The paper, thus, contributes to an emerging empirical literature on the diffusion of open source innovation. First, it introduces a dynamic model to investigate several factors that could affect the diffusion of an *open* 2-sided platform, using Mozilla as an example. Although other empirical studies in marketing have considered features of platform innovation, namely indirect network effects (e.g., Gupta et al. 1999; Basu et al. 2003), to the best of our knowledge, none has addressed features unique to the heterogeneous and interdependent diffusion of an open software platform system. Second, the paper examines the influence of a quick release strategy, a central but rarely tested assumption in OSS innovations (von Krogh and von Hippel 2006). Three, it examines the role of online, user-generated feedback on the diffusion of open innovation. Ironically, user generated content is central to the diffusion of an open platform; since the development of complementary software and the self-governing process enveloping it, generate free content that in turn could drive diffusion (Hauser et al. 2006). Finally, we extend the basic BDM to an interdependent system; and for its estimation, we innovate to design a Bayesian approach to the Extended Kalman Filter (EKF-MCMC). Our specification allows diffusion to be latent, random processes; which in turn give us the flexibility to model adoption patterns beyond the simple s-shaped ones underlying the *mixed effects*, or basic diffusion model (Goswami 2001). Similarly, we link diffusion parameters to online ratings and observational learning that varies across product and time; in this way we account for adoption heterogeneity (Dellarocas et al. 2007). Lastly, we endogenize the market potential of the complements and control for the likely dis-adoption of free software. Our results show, that observational learning and add-on ratings increase the demand for Firefox add-ons; add-ons can increase the market potential of the Firefox platform; a slow add-on review process can diminish platform success; and OSS platforms (i.e. Chrome and Firefox) compete rather than complement each other.

The main data for this study is a sample of 52 popular Firefox add-ons, free software that complement the Mozilla Firefox web browser. On the end-user side of this 2-sided platform, we collected daily reviews, user base sizes, and number of downloads for a six-year period, 2008 to 2013. For each add-on, we obtained its number of daily users, user ratings, and features of the add-on and its developer such as license types and modes of compensation for developers. Finally, for the platform, we collected data on the daily number of users of Firefox, and its main competitors, Google's Chrome and Microsoft's Internet Explorer. Though public data on daily platform use are unavailable, we were able to estimate them from other sources, including market share, and public internet global statistics.

The remainder of the paper is organized as follows. Section 2 provides the conceptual background and main hypothesis of the study. Section 3 develops the theoretical model. Section 4 presents details of the econometric model. Subsequently, section 5 presents the results of the estimation; and the implications of these results for open innovation. The paper concludes with an overview of the findings, managerial implications and limitations of the study.

*CONCEPTUAL BACKGROUND*

What then are the salient features of an OSS platform and its complements for our study? First, an OSS platform is a type of distributed innovation in which a platform owner invites third-party collaborators to develop complements (add-ons) that extend the capabilities of the platform (Sawhney et al. 2000; Kogut and Metiu 2001; Boudreau 2010). Studies report that collaborators (typically internet-based software developers and end-users) have two types of motives to participate in OSS projects: intrinsic motives, which include joy, altruism, and autonomy; and extrinsic motives, which include money, skill, and reputation (Bitzer et al. 2007; Franck and Jungwirth 2003). Moreover, the software they create remain public goods (von Krogh and von Hippel 2006), available online via distinct licenses, which govern software use and distribution. Thus, unlike that of proprietary platforms (Rochet and Tirole 2004, 2006; Parker and Van Alstyne 2005; Economides and Katsamakas 2006), the diffusion of OSS platforms arise largely from factors unrelated to price; but more likely related to OSS developer/end-user participation, interaction and OSS project governance (von Krogh and von Hippel 2006).

Thus OSS projects are managed (or governed) in ways that encourage voluntary participation; that is, the process of creating software aims to be open and transparent, decentralized and democratic (Shah 2006; O'Mahony and Ferraro 2007; O'Mahoney 2007; Markus 2007). Yet for open platform owners, software governance is a delicate balance between the ideals of democratic participation (*poetry*) and the practical task of creating good software (*pragmatism)* (Bahrami 2013; Rao et al. 2009; Krishnamurthy 2005). For example, to publish an OSS complement, a developer is often required to submit code to a review committee (O'Mahony and Ferraro 2007) of peer developers (Wang et al. 2012; Frey 2003). To make the process appear democratic, however, some platforms discourage the committee from evaluating software based upon its market usefulness, leaving that judgment to other developers and users. Similarly, OSS platforms use various forms of licenses and tools that are consistent with OS ideals but preserve some financial incentives for developers (Subramanian 2009, Nair et al. 2004, Katz and Shapiro 1994). These licenses thus may range from the very restrictive, such as General Public License (GPL), to less restrictive, such as Berkeley Software Distribution (BSD). BSD licenses allow agents to use OSS software without paying royalty on a product sold for profit, encouraging some profit motive among its community members. Alternatively, platform owners may allow its developers to solicit financial support from the end-users, or advertise a developer's online profile.

Platform owners also recognize that while democratic governance can encourage participation, it can also create chaos, arising from the numerous community feedbacks and project requests; a catalyst for this being the speed with which information moves over the internet (Feller and Fitzgerald 2000). To manage chaos, OSS developers often use a quick release system (Sharma et al. 2002), as opposed to the slow, cautious approach adopted for proprietary software. As a result, a platform may release an OSS complement version that performs only core but lacks secondary features or final aesthetics; then leverage the expertise of its community to remove bugs, improve product quality, and satisfy user requests. (See Figure 1 for Mozilla's community structure). Raymond 1999 calls this the "release early, release often" approach to software development, a central tenet of OSS that, as we note before, has been rarely tested empirically (Chen et al. 2013). Bughin et al. (2008) calls such a community interaction "co-creation," as often the same members create and use the software. Co-creation could facilitate the diffusion of an OSS platform, because it may speed creation of complements, which helps platforms compete and promote indirect network externality (Lerner and Tirole 2002).

-- Insert Figure 1 here --

Of course, the success of an OSS platform may not depend only on the software it creates, but also on the active use of its complements and the tasks the platform undertakes on behalf of its users-developers, such as responding to support requests, and writing reviews (Lerner and Tirole 2002). These activities studies suggest can build a platform's social capital, which in turn can determine a platform's diffusion (Roberts et al. 2006), where here measures of social capital include its numbers of adopters, its complements' online user ratings, and its daily user counts (Moe and Trusov 2011, Chevalier and Mayzlin 2006). Online rating valence and dispersion then act as the community's WOM, while the daily user counts could act as observational learning signals; both surrogate measures of direct network effects (Lakhani and Von Hippel 2003). Notable, meta-analysis confirms that market competition can moderate (direct and indirect) network externality on OSS platform adoption (Subramanian 2009, Nair et al. 2004, Katz and Shapiro 1994, Banaccorsi and Rossi 2003).

*HYPOTHESIS DEVELOPMENT*

Given the above background, we can now suggest the following factors, and propose how they may moderate the diffusion of an OSS platform and its free complements: i) the platform's competitors; ii) the add-on review process; iii) network externality of complements; iv) frequent releases of the platform and its complements; and v) online ratings and usage count signals.

*OSS Platform Competition*

A key distinction between OSS and proprietary platforms is the absence of a pricing mechanism in the former (e.g. Katz and Shapiro 1985, 1994; Shapiro and Varian 2013). Thus because OSS is free we may be inclined to think that users would adopt multiple platforms (e.g., *Google Chrome* and *Mozilla Firefox*); for each platform might have unique features and these are freely available (Cai et al. 2008). Consumer search theory suggests, however, that consumers may face cognitive if not monetary costs when adopting new products (Johnson et al. 2003); and we expect learning costs to be high for software. Moreover, downloading and using multiple platforms (e.g., web browsers) at the same time might diminish the quality of the end-user's experience with each platform because of the greater demand for memory and CPU. Thus, we should observe substitution rather than complementary patterns not only among proprietary platforms, but also among open platforms (Rochet and Tirole 2003). Therefore we predict:

> Hypothesis 1 (H1): As the number of end-users of an OSS (or proprietary) platform increases, the number of end-users of its peer OSS platform decreases.

*OSS Review Committee Governance*

OSS project governance is a key predictor of platform success (O'Mahony and Ferraro 2007; O'Mahony 2007). Mozilla's governance system is similar to that described above; that is, it also combines *poetry and pragmatism*[2]; poetry being Firefox's democratic, loose project governance; and pragmatism being the process through which Mozilla manages users and developers. For example, Mozilla uses a review committee of peer-developers (i.e., the Add-on Mozilla Organization or AMO) to evaluate the quality of code submitted, and to recommend which software Mozilla should be accept unto its platform. Presumably, a more proactive review committee should facilitate platform diffusion. Quick reviews and acceptance of quality software and rapid response to developers should foster greater community participation; and OSS developers are likely to generate greater positive WOM for projects when the review committee is actively contributing. In addition, faster acceptance of a software complement unto the platform would be an extrinsic and/or extrinsic reward for developers (Bitzer et al. 2007), which in turn may attract more developers to the platform. As a result, we predict:

> Hypothesis 2 (H2): As the contributions of the OSS review committee increase, the number of adopters of an OSS platform increases.

*Network Externalities*

A large installed base of platform users not only influences others to adopt the platform, it also attracts more third-party developers, who in turn may increase the number and diversity of complementary items, which then attract more adopters to the platform. What are some ways through which complementary items (i.e., add-ons) boost platform adoption? First, add-ons provide potential adopters with information that enables them to learn more about the features and utilities of the platform (Cottrell and Koput, 1998). Second, they increase the value of a platform to potential adopters by allowing them to exploit its features more fully (McIntyre and Subramaniam 2009). The translation add-on of Mozilla Firefox, for example, enhances users' ability to surf the web via access to more information in foreign languages. Third, add-ons increase the confidence among potential adopters that a platform will not disappear in the short term (Adner and Kapoor, 2010). Thus, we conclude:

> Hypothesis 3 (H3): As the number of the OSS complements increases, the potential size of the OSS platform community also increases.

---

[2] https://clarity.fm/questions/270/answers/354/share

*Release Strategy*

OSS developers issue more frequent software releases than developers of proprietary software (Bonaccorsi and Rossi 2003, Feller and Fitzgerald). A faster release frequency may attract more community contributions (e.g., bug fixes), which then help to improve software quality. A faster release would also reward contributors if it shortens the time it takes to accept their product suggestions (Raymond 1999). Lastly, a quick lease can signal the *energy* and momentum of an OSS project, and this often give users more confidence to adopt new software. Thus, we hypothesize the following:

> Hypothesis 4 (H4): As an OSS platform or its developers release new versions, adoptions of the OSS complements increase.

*Effects of End-User's Generated Contents*

As mentioned earlier, OSS end-users generate online word of mouth (WOM), and observational learning signals. Online WOM may include 1) the valence and variance of product ratings; and 2) observational learning the number of daily users of the add-on. These two signals indicate the direct network effects of the community's opinions and actions (respectively) on add-on adoption. That is, the valence gives adopters efficient access to opinions of the OSS community (Henning-Thurau and Gwinner 2004); so more positive valences should spur adoption (Chevalier and Mayzlin 2006). Conversely, the variance of the ratings reflects the community's valuation uncertainty (Sun 2012). Nevertheless, although a consumer may avoid adopting proprietary software with high valuation uncertainty, a risk-averse user may still decide to adopt an OSS complement because the expected benefit of free software may outweigh the expected loss of (say) a malicious *Trojan* (Golden 2005). Indeed, because OSS add-ons are free, a user might embrace the potential uncertainty of finding useful software, without paying for it (Martin et al 2007; Water 2012). Furthermore, a large number of users of the complement may signal its value to potential adopters, who in turn may put different weights on online ratings and observational learning. As a result, studies tend to separate the effects of observational learning from those of online WOM; for the former induces herding behavior (Chen et al. 2011); but the later signals to potential adopters the complement's the lower cognitive cost of use and user friendliness. Thus, we hypothesize the following:

> Hypothesis 5a (H5a): As the rating valence and dispersion of an OSS complement increase, the adopters of the OSS complement increase.
>
> Hypothesis 5b (H5b): As the numbers of the daily users of an OSS complement increase, the adopters of the OSS complement increase.

*MODEL DEVELOPMENT*

Our main task is to quantify the factors that can influence the interdependent diffusion of an open platform and its many add-ons. As a result, we build upon the basic Bass diffusion model, BDM (Bass 1969), given its parsimony and the intuitive interpretations of its parameters (Hauser et al. 2006). Specifically, we adopt the BDM to build system of differential equations, subsequently expressed in state space form. Thus, the first equation of the system models diffusion of the platform, and the others the diffusion of its add-ons. Our model has several notable features. For example, we allow diffusion to be latent, random processes; which in turn give us the flexibility to model adoption patterns beyond the ones underlying the *mixed effects*, or basic diffusion model (Goswami 2001). Similarly, we link diffusion parameters to online ratings and observational learning from daily use; these variable vary across product and time; and so in this way we account for adoption heterogeneity (Dellarocas et al. 2007). In addition to testing the above hypotheses, H1-H5, the model captures the role of developer incentives and license types; as well as controls for churn across add-ons. We model churn or dis-adoption because OSS complements are free but often non-unique; and consequently consumers may find them simultaneously easy to adopt and dis-adopt. Similarly, because only platform users can adopt a complement, we allow the market size of the complement to be some proportion of platform adopters, as adoption evolves over time. This proportion, our "*relevance*" parameter, should reflect the fact OSS complements often have narrow market appeal.

*Diffusion of OSS Platform*

We begin with a diffusion model for the platform that reflects the indirect externality of its complements (e.g., Gupta et al. 1999), as well as other external forces that can influence its adoption. Formally, we let $m_t$ denote the latent cumulative number of adopters of the platform at time *t*, where *M* is the market potential; and *p* and *q* the external and internal market forces' parameters, elsewhere termed innovation and contagion, respectively. Thus the diffusion, differential equation of the platform has the form:

1) $$\frac{dm_t}{dt} = (p_t + q\frac{m_t}{M_t})(M_t - m_t), t = 1,,,,,T$$

To model the influence of competition (*H1*) and governance (*H2*) on the platform's diffusion, we adopt the method proposed in Horskey and Simon (1983), modeling the external force parameter $p_t$ as the time-varying function:

2) $$p_t = p_0 + \mathbf{X}_t \mathbf{\beta} + \mathbf{Z}_t \mathbf{\rho}$$

where $p_0$ denotes the unobserved factors; $\mathbf{X}_t$ is a vector of daily usage of *Chrome* and *Internet Explorer*; $\mathbf{Z}_t$ is a vector of developer nominations and the review committee (*AMO*) 's performance; $\mathbf{\beta}$ and $\mathbf{\rho}$ are vectors of parameters to be estimated. We include Chrome and Internet Explorer as competitors of Firefox, based upon media evidence of the close competition among these three (two open source and one proprietary) browsers.

In addition, we quantify indirect externality (*H3*) by allowing the market potential of the platform to vary with the number of complements that OSS community creates on the platform (Gupta et al. 1999; Stremersch et al. 2007). In other words, new complements may attract new platform adopters by (for example) uncovering regions of unmet consumer needs. Formally, we let the platforms' market size evolves with following process:

$$3) \quad M_t = M_0 + A_t \kappa$$

where $M_0$ then denotes the unexplained market potential; $A_t$ the accumulative number of add-ons up to time *t*; and $\kappa$ a parameter to be estimated. Thus, our platform model (equation 1) explicitly captures forces consistent with indirect network externality (equation 3), and those with direct network externality, as implied by the adoption ratio, $qm_t/M_t$ (Bass 1969).

*Diffusion of OSS Complements*

Given the platform specification, we now describe the interrelated diffusion of its complements, again using the BDM framework. The novelty here is that we endogenize the market potential of the complements and control for their dis-adoption. Formally, we let $n_{jt}$ denote the latent cumulative number of adopters of the OSS complement *j* at time *t*; then diffusion of each complement becomes:

$$4) \quad \frac{dn_{jt}}{dt} = (p_{jt} + q_{jt}(1-\delta_j)\frac{n_{jt}}{\alpha_j m_t})(\alpha_j m_t - n_t) - \delta_j n_t, \; t=1,,,T_j, \; j=1,...,J$$

where $m_t$ represents the latent cumulative number of *platform adopters* at time *t*; $p_{jt}$ and $q_{jt}$ the external and internal diffusion parameters; $\alpha_j \in (0,1)$ the OSS complement's *relevance* parameter; and $\delta_j \in (0,1)$ the dis-adoption rate. That is, we use the method outlined in Libai et al. (2009) to model churn, assuming that the fraction that dis-adopts, $\delta_j$ does not spread word of mouth for the software. We think dis-adoption is a relevant feature here, given the low-costs and often non-uniqueness of OSS complements, and their analogy to services, which users can and often choose to dis-adopt (Keck 2015). Also, because many OSS complements cannot be used without adopting an OSS platform (e.g., *Firefox* add-ons) and add-ons tend to be narrowly defined, we restrict the relevance parameter, $\alpha_j$ to the interval $(0,1)$.

To study the effects of a quick release strategy on OSS (*H4*) complement diffusion, we model external or innovative factor $p_{jt}$ of each complement as the time-varying function:

5) $$p_{jt} = p_{0j} + p_{1j}PV_{jt} + p_{2j}AV_{jt}$$

where $p_{0j}$ denotes the unobservable component; $PV_{jt}$ and $AV_{jt}$ the new releases of the platform and complements, and their parameters $p_{1j}$ and $p_{2j}$, respectively. To preserve a smoothed effect of new software release as suggested in Wiggins et al. (2009), and consistent with *consumer procrastination theory*, we smooth the new releases dummy as follows:

6) $$AV_{jt} = \gamma^{t-\tau}$$

where $\tau$ is the last version release time, and $\gamma$ the decay factor. We set $\gamma$ to 0.89 the estimated decay factor of a discrete time analog model of Nerlove and Arrow (1962). We follow the same procedure to create $PV_{jt}$. Later we consider the potential endogeneity of these variables, and robustness of the smoothing procedure used to create them (See Web Appendix A).

Next we investigate the role of quality signals (*H5*) generated from the user community by modeling the imitation (WOM) factor $q_{jt}$ as the function:

7) $$q_{jt} = q_{0j} + q_{1j}RTV_{jt} + q_{2j}OL_{jt} + q_{3j}STAVG_{jt}$$

with unobserved component $q_{0j}$, and consumer learning signals -- the variance of the complements' ratings $RTV_{jt}$; the average ratings $STAVG_{jt}$; and the observational learning from the daily usage $OL_{jt}$, with parameters $q_{1j}$, $q_{2j}$ and $q_{3j}$, respectively. Here observational learning, $OL_{jt}$ is the fraction of the daily users of an OSS complement $j$ obtained from the total number of daily users of add-ons in $j$'s category (see equation 8); that is, we assume consumers not only care about absolute add-on usage numbers; but consistent with prospect theory, we also assume they compare the daily use of similar add-ons within a category (Kahneman and Tversky 1979).

8) $$OL_{jt} = \frac{Usage_{jt}}{\sum_{j \in c} Usage_{jt}}$$

Finally, to explain heterogeneity in the parameters of our joint diffusion model, we use a vector of license and potential business model information of each OSS complement. This vector includes a dummy variable for whether the add-on has the options of "ask for money contribution" or "meet the developer", and whether its license is one of the following types: "Fully Free", "Restricted", or "Mozilla". Formally, we define the vector of parameters as $\Theta_j = (\alpha_j, \delta_j, p_{0j}, p_{1j}, p_{2j}, q_{0j}, q_{1j}, q_{2j}, q_{3j})$. The hierarchical model explains this vector of parameters as follows based upon the vector of business model and license dummy variables $D_j$:

9) $$\Theta_j = D_j \eta + \varepsilon_j$$

Figure 2 gives a conceptual description of our model.

-- Insert Figure 2 here –

Thus, system of equations 1-4 is a flexible specification that reflects some key features of platform-complement diffusion. First, it captures the indirect network effect of complements on platform adoption, allowing complements to expand the potential number of platform adopters. Second, the market potential of the complement is proportional to the number of platform adopters; third, it allows for complement dis-adoption, a likely scenario for OSS complements; fourth, it captures the indirect network effect through online WOM and user activity; and finally, it allows time varying parameters in the bass model, which then makes it simple to test features of open source innovation.

*EMPIRICAL ANALYSIS*

So far we have presented a theoretical model of diffusion. To address our substantive questions (H1-H5), we will have to recast the model to an empirical form, address identification concerns, and present an estimation procedure to recover model parameters. First, however, let us consider the data available to help identify these parameters.

*Data Description*

Our main data is an unbalanced panel of daily downloads of 52 Firefox Add-ons -- free-software that extends the Mozilla browser to include features such as games and entertainment, privacy and security, language support, and web development tools. Tables 1-2 define the relevant model variables, and tables 3-6 present basic statistics for them. First, we obtained information about the daily cumulative number of Add-ons from Mozilla's website; and daily usage data for Mozilla Firefox, Google Chrome and Microsoft (MS) Internet Explorer (IE) browsers for 1686 days, from 2008 to 2013; with Firefox Add-ons launched at the different points during the period. We approximated the daily users of each web browser (i.e. Mozilla Firefox, Google Chrome and Microsoft Internet Explorer) by multiplying their daily market shares by the monthly number of computer hosts. Observing a linear pattern of growth in the internet hosts, we interleaved the daily number of hosts, as the demand or supply of personal computers is mature and hence stable.

-- Insert Tables 1-2 here –

-- Insert Tables 3-6 here --

We also obtained data on add-on/platform releases, developer reward incentives, add-on license types, and OSS governance. We smoothed the OSS new release variables using the procedure outlined earlier. Figure 3 shows the evolution of the smoothed new release measures for a sample of four add-ons. For incentives measures, we extracted data from the main page of each add-on that indicates whether its developer preferred to solicit money from or to meet end-users. In addition, Mozilla Firefox uses various licenses to protect the intellectual property of their developers (See Tables 2 and 3); thus we extracted licenses data for each add-on. Finally, to publish on the Mozilla Platform's website an add-on should pass the filter of the Add-on Mozilla Organization review (AMO) committee, a self-organized team of the experienced add-on developers who review add-ons or new releases submitted by fellow developers. We mined the monthly performance data (from Mozilla's website) of this review committee.

-- Insert Figure 3 here --

There are 19,211 online ratings in our sample (of 52 add-ons) in the period of the study. We used historical ratings data to build rating valences and rating variances, using the same process Mozilla Firefox uses to generate them for its OSS community. Observation learning comes from the visible, daily users of the add-on; on average 0.9M OSS community members use add-ons, with a variance of 2M. Lastly, we transformed several variables to avoid multicollinearity and thus improve our computations. That is, we rescaled and demeaned: AMO review committee's performance; the daily usage of the Chrome and the IE; rescaled the daily use of the Firefox and the cumulative daily downloads of the Firefox Add-ons; and demeaned product ratings mean and variance; observational learning from daily users(after same scaling); and the smoothed new releases' data.

*Empirical Model*

Admittedly, the diffusion processes and the data described earlier could be prone to process and measurement errors, in other words, a form of unobserved temporal heterogeneity. Thus first we assume the platform and complement diffusions to be latent, random processes; and consequently augment the differential equations 1-4, for $m_t$ and $n_{jt}$ with additive random, normal errors; that is, equations 1-4 now become differential equations for the means of the platform and complement diffusion processes. As mentioned earlier, treating the diffusion processes (and its parameters) as random, give us the flexibility to model other growth patterns beyond the standard s-shaped pattern (Karmeshu and Goswami 2001). We should also discretize these differential equations to obtain a model in state-space form (equations 11 and 10). This discrete form has several immediate advantages: it obviates the need to solve a large, intractable system of non-linear of differential equations (1 platform + 52 add-ons); and so is convenient for handling multivariate data (i.e., many complements) and the nonlinear dynamics embedded in the diffusion process (see, for example, Durbin and Koopman 2001, pp. 51–53). Secondly, we assume $y_{pt}$ the observed cumulative number of adopters of the OSS platform, and $y_{jt}$ those for the complement *j* at time *t* are functions of the latent adopters $m_t$ and $n_{jt}$ (respectively); and zero mean, normally distributed errors, $\varepsilon_t$ and $\varepsilon_{jt}$. So our resulting empirical model becomes:

8) $$y_{pt} = m_t + \varepsilon_t, \text{ where } \varepsilon_t \sim N(0, V_p), t = 1,...,T$$

9) $$y_{jt} = n_{jt} + \varepsilon_{jt}, \text{ where } \varepsilon_{jt} \sim N(0, V_j), j = 1,,,J; t = 1,...,T_j$$

10) $$\Delta m_t = +(p_t + q\frac{m_{t-1}}{M_t})(M_t - m_{t-1}) + \omega_t, \text{ where } \omega_t \sim N(0, W_p)$$

11) $$\Delta n_{jt} = (p_{jt} + q_{jt}(1-\delta_j)\frac{n_{jt-1}}{\alpha_j m_{t-1}})(\alpha_j m_{t-1} - n_{jt-1}) - \delta_j n_{t-1} + \omega_{jt}, \omega_{jt} \sim N(0, W_j)$$

with zero mean normal errors $\omega_t$ and $\omega_{jt}$ orthogonal to $\varepsilon_t$ and $\varepsilon_{jt}$.

*Identification and Endogeneity*

Before reviewing the estimation method we need to address potential endogeneity and identification concerns. For example, *AMO performances* and *OSS release* frequency are two potential sources of endogeneity. The release frequency may correlate with OSS adoption costs because adoption costs, which we do not observe, can spur OSS diffusion. Similarly, AMO performance can be higher due to higher levels of developer and end-user participation; and community participation could influence OSS project success (Grewal et al. 2006; Lerner et al. 2006). Thus, we test for endogeneity, employing a latent instrumental variable (LIV) approach (e.g., Naik and Tsai 2000; Rutz et. al 2010), but found no evidence of such in our sample to bias estimates of these two variables (See Web Appendix A).

Also our long time series and the variations across Firefox's add-ons allow us to identify the diffusion parameters. Moreover, our daily data should mitigate potential interval bias attributed to discrete-time bass diffusion models (Xie et al. 1997); for Putsis and Srinivasan (1999) report less severe interval bias when studies use monthly rather than the annual data. Lastly, the MCMC estimation employed, not only eases dealing with unbalanced nature of our data, but also allows us to characterize the inherent uncertainty in the diffusion parameters (Lenk and Rao1990; and Putsis and Srinivasan 1999).

*Model Estimation*

The resulting response model (8-11) is dynamic and nonlinear in the state parameters, $n_{jt}$ and $m_t$; thus we require methods appropriate to estimate their densities. Traditional methods such as OLS and MLE cannot handle non-linear dynamics and heterogeneity well, and non-linear least square (NLLS) requires analytic solutions. Thus, we thus adopt a Bayesian Sequential Monte-Carlo (SMC) approach given the flexibility of these algorithms for estimating non-linear, dynamic systems. Moreover, given the above discrete model and the normal assumptions, two SMC algorithms emerge: the Extended Kalman Filter (EKF) and the Unscented Kalman Filters (UKF), both methods work with linear-normal approximations. But the EKF would approximate the non-linear platform and add-on diffusion equation (10-11), while the UKF instead approximates the posteriors of the state variables by sets of sample points (Ristic et al. 2004). Thus, for highly nonlinear Gaussian systems the UKF produces typically better results, but our diffusion model is quadratic and so we adopt the EKF-MCMC, our Markov Chain Monte-Carlo approach to the EKF, which subsequently performs well. Also, in our application the EKF is more computationally efficient given the sizeable dimension of our state space (52 add-ons) and the maximum length our time-series (T=1686).

*Extended Kalman Filtering and MCMC (EKF-MCMC) Algorithm*

Our estimation task then is to recover the following components for the platform and each of its complements $j$: time-varying (state space) components $m_{0:T} = \{m_t\}_{t=0}^{T}$ and $n_{j,0:T} = \{n_{jt}\}_{t=0}^{T_j}$; regression components for H1-H5, $\Omega_p = \{p_0, \beta, \rho, M, \kappa\}$ and $\Omega_j = \{\{p_{ij}\}_{i=0}^{2}, \{q_{ij}\}_{i=0}^{3}\}$; and variance components, $\{V_i, W_i, i \in (j,p)\}$ ; all defined in equations 1-11. An EKF-MCMC strategy would thus seek to simulate the joint posterior of all parameters, given platform $y_{p,1:T} = \{y_{p1}, y_{p2}, ..y_{pT}\}$ and complement $y_{j,1:T_j} = \{y_{j1}, y_{j2}, ..y_{jT_j}\}, j = 1,..,J$ adoption observations. It can accomplish this by indirectly sampling iteratively from a sequence of conditional posteriors: $p(m_{0:T}|y_{p,1:T}, \Omega_p, V_p, W_p) \leftrightarrow p(\Omega_p, V_p, W_p|m_{0:T})$ and $p(n_{i,0:T}|y_{j,1:T}, \Omega_j, V_j, W_j) \leftrightarrow p(\Omega_j, V_j, W_j|n_{i,0:T})$. Notably, conditional on the time varying parameters $m_{0:T}$ and $n_{j,0:T}$ one can sample the regression and variance components using standard MCMC ideas (See Rossi et al. 2005). The conditional posteriors, $p(m_{0:T}|..)$ and $p(n_{i,0:T}|...)$, however, are unavailable analytically since the BDMs are nonlinear in their respective time-varying parameters; hence the standard Kalman filter algorithm does not apply. We thus apply the Extended Kalman Filter, an approximate filtering approach based on the Taylor series expansion of the non-linear bass equations. For example, for the platform model the approximation gives:

11) $\qquad m_t = g(a_{t-1}) + J_t(m_{t-1} - a_{t-1}) + \omega_t$

where the mean function $g(m_{t-1}) = m_{t-1} = (p_t + q\frac{m_{t-1}}{M_t})(M_t - m_{t-1})$ and the Jacobian $J_t = \left[\frac{\partial g(m)}{\partial m'}\right]_{m=a_{t-1}}$ are evaluated at mean $a_{t-1} = E(m_{t-1}|D_{t-1})$. We linearize the diffusion model for each complement in a similar way. With this linearization, we can sample $p(m_{0:T}|..)$ and $p(n_{i,0:T}|...)$ using the standard FFBS algorithm (Fruhwirth-Schnatter, 1994; Bass et al. 2007).

*ESTIMATION RESULTS*

Tables 7-11 and figures 4-6 summarize the main findings of our study, which includes the evaluations of a) hypotheses H1-H5; and b) the robustness and fit of the proposed model. First, table 7 compares the performance of our proposed model against those of nine alternatives in terms deviance information criteria (DIC). Notably, the proposed model outperforms all alternatives in terms of DIC, a suitable criterion for hierarchical model selection (Spiegelhalter et al. 2002). Similarly, Figures 4 and 5 reports the 1-step-ahead forecast the Firefox platform and four of its complements; all track the data with some precision as shown in the mean absolute deviation (MAD) and mean squared error (MSE) scores in Table 8. Thus these fit and forecast measures point to the ability of our proposed model to capture the underlying diffusion process.

Table 9 reports estimates for the Firefox platform diffusion model. First, the effects of both open $\rho_1$ and proprietary browsers $\rho_2$, *Microsoft IE* and *Google Chrome* (respectively), on the diffusion of Firefox are negative and significant (H1). This is consistent with media conjecture that both types (OSS and proprietary platforms) compete for end-users (e.g., in the so-called "Browser Wars," See Bott 2014); but still mildly surprising given that we could easily imagine OSS end-users choosing both Chrome Firefox. With regard to platform governance, we find the effect of the AMO review process to be positive and significant (H2), which again suggest the importance of the OSS community, and its self-governance to platform success. We will return to this issue later. Finally, the effect of the cumulative number of add-ons on the marketing potential of the Mozilla platform $\kappa$ is positive and significant (H3); this suggests one mechanism for the effect of indirect network externality on an OSS platform, and underscores the importance of add-ons to platform diffusion. This finding is even more interesting because the direct network effect parameter $q_0$ is insignificant, which implies that the appeal of Firefox has more to do with the availability of it complements.

Table 10 shows estimates for the OSS add-on diffusion models. The effects of new releases (add-ons $p_1$ and platform $p_2$) have positive and significant effects on the diffusion of the Firefox's add-ons. This seems consistent with a basic but largely untested tenet of OSS ideology (Raymond 1999), that releasing early and frequently help speeds OSS project success and now diffusion (H4); a prescription that runs counter to proprietary software development practices. The effects of quality signals -- rating valence, rating variance (H5a) and observational learning (H5b) -- on OSS complement diffusion are all positive and significant. Marketing studies on proprietary goods have found similar results for the effects of rating valence (Chevalier and Mayzlin 2006) and observational learning (Chen et al. 2011) on product performance. Our finding of a positive effect of rating variance for OSS (non-monetized software) diffusion stands in contrast to those of Sun (2012), who found a negative effect for the proprietary goods. Note that both Martin et al. 2007 and Clemons et al. 2006 found a similar positive effect of dispersion on the performances of movie and beer brands, which suggests that consumer behaviors in these two categories may share some similarity with those in a OSS community. Nevertheless, the founding advocate of free software, *Richard Stallman* once said in reference to OSS: "This is a matter of freedom, not price, so think of "free speech," not "free beer."[3] We thus surmise that our finding of a positive impact of variance on OSS diffusion maybe be attributed more to the joy of experiencing freedom, and less to the joy of surprise. Finally, Table 10 reports small, positive estimates of churn $\delta_j$ and the relevance $\alpha_j$ parameter. The histogram in figure 6 shows a left skewed distribution for the relevance parameters, which suggests that in an OSS context an innovation does not have to be highly relevant to obtain large numbers of potential adopters; so the OSS ecosystem allows people with heterogeneous preferences to co-exist. We find also that the daily churn of the OSS complements are on average 1.74%, which is closer to monthly estimates for online brokerage, books, and satellite radio in Libai et al. 2009 than to the yearly estimates for cellular phones.

---

[3] http://www.gnu.org/philosophy/open-source-misses-the-point.html

Lastly, table 11 shows estimates that attempt to explain heterogeneity in add-on parameters as a function of license types, and OSS developer incentives. Notably, we find no significant effects of license or incentive types on add-on diffusion, though studies suggest these could affect the success of OSS innovation (Lerner and Tirole 2002). Yet our null findings in the case of Mozilla may still be reasonable. For example, at this point in our sample, Mozilla is a successful and stable OSS platform, and so these seemingly extrinsic incentives may not influence the level and quality of developer participation or signal OSS add-on quality to end users. Moreover, many software complements rarely have economic value to developers, who would freely contribute them to the platform without the need to protect their intellectual property, and so here licenses restrictiveness becomes less relevant. For these reasons, we are not entirely surprised by these findings.

*AMO EDITORIAL EFFORT REALLOCATION*

We have seen that OSS governance, the AMO editor contributions, can improve platform success (*H2*). As a result, though Mozilla may not have direct control of the third-party AMO; Mozilla could be tempted to use soft power to organize events[4] or hire temporary staff to influence the pace of the editorial process, to the satisfaction of the add-on developer community. We consider the result of one such scenario. That is, we consider whether Mozilla could have allocated the editorial effort differently to increase platform diffusion; and then compare the features of the actual AMO effort to one that Mozilla would have implemented. For this scenario, we assume the AMO's total effort is fixed, and that Mozilla internalizes the editorial process in order to maximize the diffusion of Firefox. Our model task then is to recover a best editorial committee effort over 1,424 days of the study.

---

[4] https://blog.mozilla.org/addons/2009/06/29/amo-review-queue-burndown-a-huge-success/

For this purpose we solve a large scale non-linear problem to find the optimal daily editorial effort of AMO editors. The problem can be conceived as finding a sequence of the proposed contributions $\{pc_t\}_{t=1}^{T}$ that create the largest total cumulative diffusion. That is, we have:

P1)
$$\max_{pc_1,...,pc_T} \sum_{t=1}^{T} E[y_t | D_{t-1}]$$
$$s.t. \sum_{t=1}^{T} pc_t \leq B_{AMO}, t = 1,...,T$$

where $E[y_t | D_{t-1}]$ is the daily 1-step ahead diffusion forecast and $B_{AMO}$ the current total contribution of the Firefox AMO editors. At each month Mozilla Firefox decides on the level of contribution; hence, there are $K^T$ effort levels, where $K$ denotes the sum of actual AMO observed effort over the T days of the planning horizon. The solution to this problem involves searching over a set of admissible effort sequences of $\{pc_t\}_{t=1}^{T}$ to find a sequence that yields the largest level of the expected total diffusion, given the estimated non-state parameter values for the dynamic model. The total expected diffusion is based on the information available at the beginning of the planning stage, so we use an Extended Kalman Filter (forward filtering step), given the non-state parameter estimates to maximize the cumulative one-step ahead forecast of the diffusion path $\{\hat{Y}_t\}_{t=1}^{T}$. Now because finding the optimal solution requires exponential time order processing, we use the genetic algorithm (GA) available in MATLAB. Recall, the GA starts with a random candidate population; and through the processes of selection, cross over and mutation, it produces better off-springs (schedules) in subsequent generations and converges on a set that is most likely to contain the optimal schedule.

Table 12 reports the actual next to the optimal or model-based AMO contributions obtained from our solution to P1. These model-based contributions would generate an additional 15 million users of the Mozilla platform. Moreover, the solution provides two intuitive results. First, simple calculations show that the cumulative contributions over time derived from P1 are higher in nearly all cases; and second, the contributions are more stable than the actual contributions (Standard deviation: 252.98 vs 441.56). That means Mozilla could have probably increased the diffusion of *Firefox* by increasing the pace but reducing the fluctuations (or variance) of its editorial response to nominations from the developer community. These specific actions seem consistent with case studies on Mozilla Firefox (Rao et al. 2009), which view the developer community as critical to Mozilla's long-term success against Internet Explorer, (at the time) the default platform for many end-users.

-- Insert Tables 12 here –

*CONCLUSION AND LIMITATIONS*

We study the interdependent diffusion of an open source software (OSS) platform and its software complements, new product diffusion in the absence of typical marketing incentives. Here an OSS platform is a type of distributed innovation in which an owner invites third-party collaborators to create add-ons/plug-ins that extend the features of the platform. Our goal was to quantify the effects of *i*) platform competition;  *ii*) add-on reviews or OSS governance; *iii*) network externality of complements; *iv*) frequent software releases; and *v*) online word of mouth and observational learning signals, all on OSS diffusion.

To measure these effects we extend the Bass model (BDM) to allow for parameter dynamics, heterogeneity (e.g., *p*, *q*), endogeneity (e.g., market potential, *M*). We apply the extended model to a canonical example, the diffusion of the *Mozilla Firefox* browser and 52 of its add-ons. We re-cast the resulting 53 differential equations into non-linear, discrete-time, state-space forms; and then estimate this interdependent system using an MCMC approach to the Extended Kalman Filter (EKF), a Sequential Monte Carlo method that works with normal-linear approximations. Our EKF-MCMC algorithm is preferred here to other Gaussian filters such as the Unscented Kalman Filter (UKF) and the Augmented Kalman Filter with continuous state space and discrete observations (AKF-CD) given the low order of our non-linearity but high dimensions of our state-space; the lengths of our sample (T=1686 days) and the importance of heterogeneity. Moreover our daily data mitigates potential interval bias attributed to discrete-time bass diffusion models.

We obtained several interesting results. First, results show the *indirect network* effects (of add-ons) on the platform to be positive and significant; but the *direct network* effects to be insignificant. Thus, OSS platforms should also encourage content creation (i.e., add-ons) rather than focus exclusively on generating platform awareness. Results also show that the add-on (editorial) review process can help spur platform diffusion. As a result, a platform could be tempted to internalize the editorial review process or use more soft power to influence it. In fact, our simulation showed that Mozilla could probably increase its diffusion by reducing the fluctuations of and speeding-up the AMO's response to *nominations* from its developers. Similarly, ratings valence and variance, and observational learning all had positive effects on add-on diffusion. Platforms owner should therefore make these quality signals readily transparent because they help effect the match between the two sides of the platform, developers and end-users.

Finally, some of our findings point to the delicate balance OSS platforms must maintain between *i*) the ideals of democratic participation (e.g., accepting add-ons without screening for market needs) (*poetry*); and *ii*) the practical task of creating competitive software (*pragmatism)*. For example, we find that OSS platforms compete rather than complement each other and so

OSS platforms may need to consider marketing strategies to differentiate their products; ideas that traditionally motivate proprietary software teams; but could alienate OSS developers. Moreover, our findings of the positive influences of the "release early and release often" concept of Raymond (1999) gives the OSS platforms a tool to increase societal welfare by providing more innovations faster, but also to compete with open or even proprietary systems. Yet we find that OSS license types and commercial incentives for developers had no effect on platforms performance; this means OSS platform could be more democratic (than proprietary) to allow individuals with broad types of incentives to participate in OSS innovation, given that they comply with OSS rules.

Nevertheless, our work has several limitations that potentially could be addressed with the benefit of additional data. First, with individual level data on the activities of OSS members, we could better explain the macro level findings above. In this case our macro-level diffusion models would have to emerge from individual behaviors of and interactions among OSS adopters. Still, we would need to invent a fruitful way to incorporate micro level data into our mixed influence, aggregate diffusion models. Moreover, for an accurate forecast of the platform and complement adoption, the models would require data on all potential adopters. Second, our study lacks information on the software code or bug feedbacks from the AMO editors to the developers; having that data may give us some knowledge on the quality or effectiveness of the OSS developer community. It would also have been useful to model the diffusion of both OSS and proprietary platforms; so we could contrast the differences in factors that drive their diffusions. Finally, even though our results are consistent with predictions, we obtained our results for an open browser platform (*Firefox*); and so it would be useful to extend our analysis to other forms of OSS platforms (e.g., OS).

Table 1: VARIABLE DESCRIPTIONS

| Variable | Description |
|---|---|
| Add-on Daily Download($y_{jt}$) | The observed cumulative of number of users who downloaded an add-on in a given day. |
| Mozilla Firefox daily Users ($y_{pt}$) | The daily number of users of Mozilla Firefox's platform. We recovered this information by multiplying the daily Mozilla Firefox market share by the monthly number of internet hosts. |
| Total number of add-ons created per day($A_t$) | The cumulative number of add-ons created by the community of developers from the inception of our data. |
| Google Chrome daily Users ($Z_t^1$) | The daily number of users of Chrome. We recovered this information by multiplying daily Google Chrome's market share by the monthly total number of internet hosts. We use this variable to explain the external diffusion market force. |
| Microsoft Internet Explorer (IE) daily Users ($Z_t^2$) | The daily number of users of Internet Explorer. We recovered this information by multiplying daily Microsoft Internet Explorer's market share by monthly total number of internet hosts. We use this variable to explain the external diffusion market forces. |
| New Version of add-on($AV_{jt}$) | An indicator variable that shows a new version is issued. We smooth it based on demand and release trend curve presented by Wiggins et al. (2009)[5] and based on consumer procrastination theory with a 0.8 factor that is the decay parameter of our estimated Nerlov and Arrow model. |
| Ask for money contribution ($M_t^1$) | An indicator variable. Some add-on pages suggest that if a user has enjoyed the add-on, they can help support the add-ons' continued development by making a small money contribution with a click of a button. |
| Meet the developer option ($M_i^2$) | An indicator variable. Some add-ons give users the option to meet the developer, to know why the add-on is created and what's next for the add-on. By clicking on the link one can see the contact information of the developer and his/her profile. |
| Fully Free License ($M_i^3$) | This is an indicator variable specifying whether the license of an add-on is either BSD or MIT/X11 License. |
| Restricted Licenses ($M_i^4$) | This is an indicator variable specifying whether the license of an add-on is either GNU or Custom License. |
| Mozilla License ($M_i^5$) | This is an indicator variable specifying whether the license of an add-on is a Mozilla License. |
| Total length of the monthly AMO nomination queue($Z_t^4$) | After nomination, the add-on status page will indicate the status of "In Sandbox: Public Nomination". This means the add-on is in the nomination review queue[6], the size of which we capture in this variable. |
| Total number of monthly AMO | In order for an add-on version to become public and readily |

---

[5] Wiggins, Andrea, James Howison, Kevin Crowston. "Heartbeat: measuring active user base and potential user interest in FLOSS projects." *Open Source Ecosystems: Diverse Communities Interacting*. Springer Berlin Heidelberg, 2009. 94-104.
[6] https://blog.mozilla.org/addons/2010/02/15/the-add-on-review-process-and-you/

| | |
|---|---|
| Editor's contribution($Z_t^3$) | available to all, it needs to be submitted to a review committee (AMO). This process is called nomination[7]. AMO engages in four types of activities:[8] full review nominations, full review updates, preliminary reviews, and response to info request. [9] Mozilla Firefox forum calls the sum of the number of responses of AMO to these incidences or nominations, total editor contributions, which we capture in this variable.[10] |
| Add-on Daily Users($OL_{jt}$) | The daily users of add-on in the previous day, visible to consumers. |
| Rating Valence Mean ($STAVG_{jt}$) | The discrete number of stars that show rate of the product out of 5. We recovered this information from the historical data of rating, the same way Mozilla Firefox generates it. |
| Rating Variance ($RTV_{jt}$) | The variance of distribution of product ratings. We recovered this from the historical data on distribution of product ratings. |

Table 2: LICENSE DESCRIPTION[11]

| Item | BSD | MIT/X11 | Mozilla | GNU | Custom |
|---|---|---|---|---|---|
| Provides copyright protection | TRUE | TRUE | TRUE | TRUE | TRUE |
| Can be used in commercial applications | TRUE | TRUE | TRUE | TRUE | FALSE |
| Bug fixes / extensions must be released to the public domain | FALSE | FALSE | TRUE | TRUE | TRUE |
| Provides an explicit patent license | FALSE | FALSE | TRUE | FALSE | FALSE |
| Can be used in proprietary (closed source) applications | TRUE | TRUE | TRUE | FALSE | FALSE |

Table 3: DESCRIPTIVESTATISTICS OF LICENSES AND INCENTIVES

| Item | Type | Frequency |
|---|---|---|
| License | Fully free (MIT/X11,BSD) | 5 |
| | Restricted (GNU, Custom) | 41 |
| | Mozilla | 2 |
| Incentive | Contribute | 25 |
| | Meet Developer | 10 |

---

[7] https://blog.mozilla.org/addons/2010/02/15/the-add-on-review-process-and-you/
[8] https://blog.mozilla.org/addons/2011/02/04/overview-amo-review-process/
[9] When someone submits a new add-on, it will have to choose between 2 review tracks: Full Review and Preliminary review; the first one checks whether the add-on is safe to use, respects user's privacy and choice, doesn't conflict with other add-ons or break existing Firefox features, is easy to use, and is worth publishing to a general audience. The second one, only requires add-on to be safe to use. Add-on with preliminary review approval appear on the site as Experimental, cant' be featured and get lower search ranking. If an add-on approved in the preliminary review track, it can be nominated to the Full Review track after a 10 day waiting period.
[10] https://forums.mozilla.org/addons/viewtopic.php?f=21&t=14313
[11] http://www.codeproject.com/info/Licenses.aspx

Table 4: ADD-ON BASIC STATISTICS

|  | Mean | SD | Min | Max |
|---|---|---|---|---|
| Daily Downloads (K) | 7.64 | 18.50 | 11.79 | 283.44 |
| Daily Users (M) | 0.88 | 1.95 | 1E-06 | 16.97 |
| Rating Valence Mean | 4.28 | 0.50 | 1.00 | 5.00 |
| Rating Variance | 1.46 | 0.76 | 0.48 | 4.20 |
| New Version of add-on indicator | 0.02 | 0.12 | 0.00 | 1.00 |
| Length of time series | 1321.90 | 456.60 | 260.00 | 1686.00 |

Table 5: PLATFORM BASIC STATISTICS

|  | Mean | SD | Min | Max |
|---|---|---|---|---|
| Mozilla Firefox daily Users (M) | 229 | 16 | 185 | 262 |
| Total number of add-ons create per day | 128 | 192 | 4 | 2,418 |
| Google Chrome daily Users (M) | 189 | 119 | 20 | 432 |
| Microsoft Internet Explorer (IE) daily Users (M) | 354 | 47 | 240 | 437 |
| Total number of monthly AMO Editor's contribution | 1,444 | 442 | 794 | 2,620 |
| Total length of the monthly AMO nomination queue | 362 | 220 | 80 | 949 |

Table 6: ADD-ON CATEGORIES BASIC STATISTICS

| Add-on Categories | Representation in our sample |
|---|---|
| Appearance | 9 (17%) |
| Bookmarks | 3 (6%) |
| Download Management | 3 (6%) |
| Photo and Multimedia | 9 (17%) |
| Game and Entertainment | 3 (6%) |
| Privacy and Security | 6 (12%) |
| Language Support | 7 (13%) |
| Alerts Updates | 6 (12%) |
| Web Development | 13 (25%) |

Table 7: MODEL COMPARISONS

| Model | Description | DIC | $p_D$ | $LL(\bar{\theta})$ |
|---|---|---|---|---|
| 1 | No Churn | 393,181,425 | 196,887,137 | 296,425 |
| 2 | No Version Carry Over | 393,249,002 | 196,921,381 | 296,881 |
| 3 | No AMO effect on platform | 393,258,460 | 196,926,790 | 297,560 |
| 4 | Interaction model ($PV_{jt} \times AV_{jt}$, $RTV_{jt} \times OL_{jt}$, $RTV_{jt} \times STAVG_{jt}$) | 393,224,624 | 196,921,662 | 309,350 |
| 5 | Unexplained internal market force of add-ons | 393,183,354 | 196,889,620 | 297,942 |
| 6 | Unexplained external market force of add-ons | 393,229,107 | 196,907,050 | 292,496 |
| 7 | Unexplained churn | 393,265,094 | 196,298,559 | 297,309 |
| 8 | No cumulative effect of add-on creation on platform | 393,220,649 | 196,903,764 | 293,439 |
| 9 | Unexplained relevance factor | 393,338,158 | 196,971,387 | 302,309 |
| 10 | Proposed Model | 393,029,826 | 196,694,177 | 179,264 |

Table 8: PERFORMANCE OF THE PROPOSED MODEL FOR FOUR ADD-ONS AND PLATFORM

| Description | MAD | MSE |
|---|---|---|
| Firefox Platform | 1.20e-04 | 2.04e-05 |
| Auto-Pager Add-on | 0.0016 | 4.71e-06 |
| Google Translator for Firefox Add-on | 0.0012 | 3.34e-06 |
| Ad-block Plus Add-on | 0.0049 | 3.97e-05 |
| Stealthy Add-on | 0.0032 | 1.32e-05 |

Table 9: PARAMETER ESTIMATES -- PLATFORM

| | Estimate | Std. Dev. | 2.5th | 97.5th |
|---|---|---|---|---|
| Market Size[12]: | | | | |
| Intercept of Market size $M_0$ | 1.54E-02 | 1.52E-06 | 1.54E-02 | 1.54E-02 |
| Total Add-ons Created $\kappa$ | 3.60E-02 | 1.51E-06 | 3.60E-02 | 3.60E-02 |
| External Market Force: | | | | |
| Unobserved external Market Force $p_0$ | 1.76E-03 | 1.52E-06 | 1.76E-03 | 1.76E-03 |
| Google Chrome competitor $\rho_1$ | -4.91E-05 | 1.51E-06 | -5.17E-05 | -4.73E-05 |
| Microsoft Internet Explorer competitor $\rho_2$ | -5.66E-04 | 1.52E-06 | -5.68E-04 | -5.64E-04 |
| AMO Total number of contributions $\rho_3$ | 3.42E-05 | 1.51E-06 | 3.16E-05 | 3.61E-05 |
| AMO Length of the Queue of nominations $\rho_4$ | 3.52E-05 | 1.51E-06 | 3.26E-05 | 3.70E-05 |
| Internal Market Force: | | | | |
| Unobserved Internal Market Force $q_0$ | 1.27E-08 | 9.02E-09 | -2.03E-09 | 2.77E-08 |
| Variances: | | | | |
| Observation Equation $v_p$ | 1.44E-02 | 4.30E-03 | 8.52E-03 | 2.30E-02 |
| State Equation $w_p$ | 1.12E-01 | 8.35E-03 | 9.75E-02 | 1.25E-01 |

---

[12] Market size is variable with time according to number of new add-ons $M_t = M_0 + A_t \kappa$

Table 10: PARAMETER ESTIMATES – ADD-ONS

|  | Estimate | Std. Dev. | 2.5$^{th}$ | 97.5$^{th}$ |
|---|---|---|---|---|
| Relevance factor $\alpha_j$ | 0.0142 | 0.0019 | 0.0104 | 0.0179 |
| Churn factor $\delta_j$ | 0.0174 | 0.0021 | 0.0132 | 0.0215 |
| External Market Force: | | | | |
| Unobserved $p_0$ | 0.0087 | 0.0018 | 0.0051 | 0.0123 |
| Add-on New Version $p_1$ | 0.0047 | 0.0011 | 0.0026 | 0.0067 |
| Platform New Version $p_2$ | 0.0059 | 0.0012 | 0.0035 | 0.0083 |
| Internal Market Force: | | | | |
| Unobserved $q_0$ | 0.0057 | 0.0014 | 0.0030 | 0.0085 |
| Rating Variance $q_1$ | 0.0131 | 0.0018 | 0.0096 | 0.0167 |
| Observational Learning $q_2$ | 0.0054 | 0.0016 | 0.0022 | 0.0086 |
| Rating valence mean $q_3$ | 0.0043 | 0.0014 | 0.0016 | 0.0070 |
| Variance: | | | | |
| Observation Equation $v_j$ | 0.0002 | 1.56E-05 | 0.0002 | 0.0002 |
| State Equation $w_j$ | 0.0002 | 1.74E-05 | 0.0002 | 0.0003 |

Table 11: PARAMETER HETEROGENEITY

|  |  | Estimate | STD | 2.5$^{th}$ | 97.5$^{th}$ |
|---|---|---|---|---|---|
| Relevance factor $\alpha_i$ | Intercept | 0.014 | 0.007 | 0.002 | 0.026 |
|  | Ask for money contribution | 0.010 | 0.016 | -0.017 | 0.038 |
|  | Meet the developer option | -0.003 | 0.020 | -0.037 | 0.030 |
|  | Fully Free License | 0.001 | 0.034 | -0.056 | 0.057 |
|  | Restricted Licenses | 0.006 | 0.027 | -0.039 | 0.051 |
|  | Mozilla License | -0.001 | 0.046 | -0.076 | 0.073 |
| Churn factor $\delta$ | Intercept | 0.014 | 0.007 | 0.003 | 0.026 |
|  | Ask for money contribution | 0.010 | 0.016 | -0.017 | 0.037 |
|  | Meet the developer option | -0.003 | 0.021 | -0.037 | 0.032 |
|  | Fully Free License | 0.001 | 0.034 | -0.055 | 0.056 |
|  | Restricted Licenses | 0.007 | 0.026 | -0.037 | 0.051 |
|  | Mozilla License | 0.000 | 0.045 | -0.073 | 0.073 |
| *External Market Force* | | | | | |
| Unobserved $p_0$ | Intercept | 0.014 | 0.007 | 0.002 | 0.025 |
|  | Ask for money contribution | 0.011 | 0.016 | -0.016 | 0.038 |
|  | Meet the developer option | -0.003 | 0.020 | -0.036 | 0.031 |
|  | Fully Free License | 0.001 | 0.034 | -0.055 | 0.056 |
|  | Restricted Licenses | 0.007 | 0.027 | -0.036 | 0.051 |
|  | Mozilla License | -0.001 | 0.045 | -0.075 | 0.072 |
| Add-on New Version $p_1$ | Intercept | 0.009 | 0.007 | -0.003 | 0.021 |
|  | Ask for money contribution | 0.000 | 0.017 | -0.028 | 0.027 |
|  | Meet the developer option | 0.009 | 0.022 | -0.027 | 0.045 |
|  | Fully Free License | -0.019 | 0.036 | -0.077 | 0.039 |
|  | Restricted Licenses | -0.005 | 0.028 | -0.051 | 0.041 |
| Platform New Version $p_2$ | Mozilla License | -0.013 | 0.047 | -0.091 | 0.063 |
| | Intercept | 0.005 | 0.007 | -0.007 | 0.016 |
| | Ask for money contribution | -0.001 | 0.017 | -0.030 | 0.028 |
| | Meet the developer option | 0.014 | 0.023 | -0.023 | 0.051 |
| | Fully Free License | -0.014 | 0.036 | -0.074 | 0.045 |
| | Restricted Licenses | -0.004 | 0.028 | -0.050 | 0.043 |
| | Mozilla License | 0.000 | 0.048 | -0.080 | 0.078 |

*Internal Market Force*

| | | | | | |
|---|---|---|---|---|---|
| Unobserved $q_0$ | Intercept | 0.014 | 0.007 | 0.003 | 0.026 |
| | Fully Free License | -0.001 | 0.034 | -0.058 | 0.056 |
| | Restricted Licenses | 0.006 | 0.027 | -0.039 | 0.050 |
| | Mozilla License | -0.002 | 0.045 | -0.077 | 0.072 |
| | Ask for money contribution | 0.000 | 0.029 | -0.048 | 0.047 |
| | Meet the developer option | -0.003 | 0.021 | -0.037 | 0.032 |
| | Asked contribution amount | 0.002 | 0.004 | -0.004 | 0.007 |
| Rating Variance $q_1$ | Intercept | 0.009 | 0.007 | -0.004 | 0.021 |
| | Fully Free License | -0.018 | 0.036 | -0.076 | 0.040 |
| | Restricted Licenses | -0.005 | 0.028 | -0.051 | 0.042 |
| | Mozilla License | -0.014 | 0.046 | -0.091 | 0.062 |
| | Ask for money contribution | 0.001 | 0.029 | -0.048 | 0.049 |
| | Meet the developer option | 0.009 | 0.022 | -0.026 | 0.045 |
| | Asked contribution amount | 0.000 | 0.004 | -0.006 | 0.006 |
| Observational Learning $q_2$ | Intercept | 0.005 | 0.007 | -0.007 | 0.017 |
| | Fully Free License | -0.015 | 0.037 | -0.076 | 0.046 |
| | Restricted Licenses | -0.004 | 0.029 | -0.051 | 0.043 |
| | Mozilla License | -0.003 | 0.048 | -0.081 | 0.076 |
| | Ask for money contribution | -0.003 | 0.029 | -0.051 | 0.046 |
| | Meet the developer option | 0.013 | 0.022 | -0.024 | 0.049 |
| | Asked contribution amount | 0.000 | 0.004 | -0.006 | 0.006 |
| Rating mean $q_3$ | Intercept | 0.006 | 0.007 | -0.006 | 0.018 |
| | Fully Free License | -0.010 | 0.036 | -0.069 | 0.049 |
| | Restricted Licenses | -0.003 | 0.029 | -0.050 | 0.044 |
| | Mozilla License | 0.021 | 0.048 | -0.058 | 0.100 |
| | Ask for money contribution | -0.001 | 0.029 | -0.050 | 0.048 |
| | Meet the developer option | 0.010 | 0.023 | -0.028 | 0.048 |
| | Asked contribution amount | 0.000 | 0.004 | -0.006 | 0.006 |

Table 12: MONTHLY EDITORIAL EFFORT -- ACTUAL VS. MODEL-BASED

| Year | Month | Actual Contributions | Model-Based Contributions |
|---|---|---|---|
| 2009 | 7 | 972 | 958 |
| 2009 | 8 | 994 | 1526 |
| 2009 | 9 | 1011 | 1642 |
| 2009 | 10 | 1375 | 1380 |
| 2009 | 11 | 794 | 1450 |
| 2009 | 12 | 1673 | 1298 |
| 2010 | 1 | 1402 | 1651 |
| 2010 | 2 | 1537 | 1201 |
| 2010 | 3 | 1159 | 1571 |
| 2010 | 4 | 965 | 1508 |
| 2010 | 5 | 903 | 1435 |
| 2010 | 6 | 835 | 1409 |
| 2010 | 7 | 1047 | 1540 |
| 2010 | 8 | 923 | 1280 |
| 2010 | 9 | 801 | 1332 |
| 2010 | 10 | 1233 | 1554 |
| 2010 | 11 | 841 | 1556 |
| 2010 | 12 | 2416 | 1455 |
| 2011 | 1 | 2620 | 1210 |
| 2011 | 2 | 1776 | 956 |
| 2011 | 3 | 1235 | 1718 |
| 2011 | 4 | 1624 | 1668 |
| 2011 | 5 | 2054 | 1523 |
| 2011 | 6 | 1834 | 1203 |
| 2011 | 7 | 1729 | 1704 |
| 2011 | 8 | 1276 | 1439 |
| 2011 | 9 | 1498 | 1594 |
| 2011 | 10 | 1346 | 1479 |
| 2011 | 11 | 1790 | 1432 |
| 2011 | 12 | 2275 | 1837 |
| 2012 | 1 | 2294 | 1345 |
| 2012 | 2 | 1360 | 1285 |
| 2012 | 3 | 951 | 1670 |
| 2012 | 4 | 1359 | 1347 |
| 2012 | 5 | 1261 | 1472 |
| 2012 | 6 | 1409 | 1407 |
| 2012 | 7 | 1391 | 1485 |
| 2012 | 8 | 1222 | 1631 |
| 2012 | 9 | 1805 | 1532 |
| 2012 | 10 | 1316 | 1538 |
| 2012 | 11 | 1720 | 1283 |
| 2012 | 12 | 1412 | 1329 |
| 2013 | 1 | 1651 | 1640 |
| 2013 | 2 | 1723 | 1492 |
| 2013 | 3 | 1470 | 1608 |
| 2013 | 4 | 1351 | 1462 |
| 2013 | 5 | 2093 | 1380 |
| 2013 | 6 | 1283 | 210 |

**Figure 1: Mozilla Community Structure**

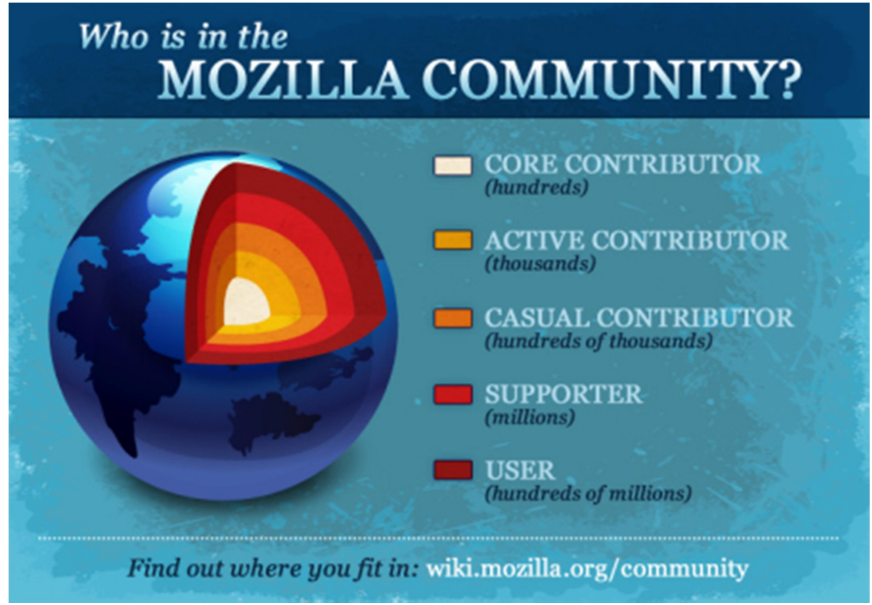

**Figure 2: Conceptual Model**

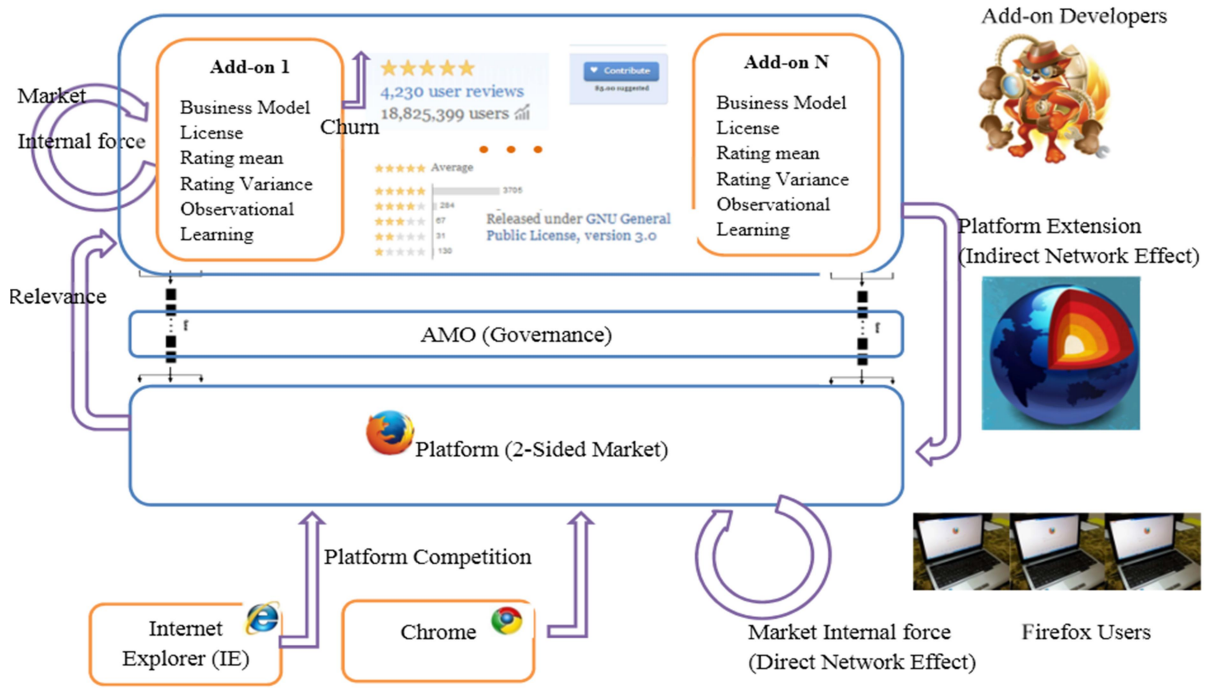

**Figure 3: Add-on Daily New Version Dynamic**

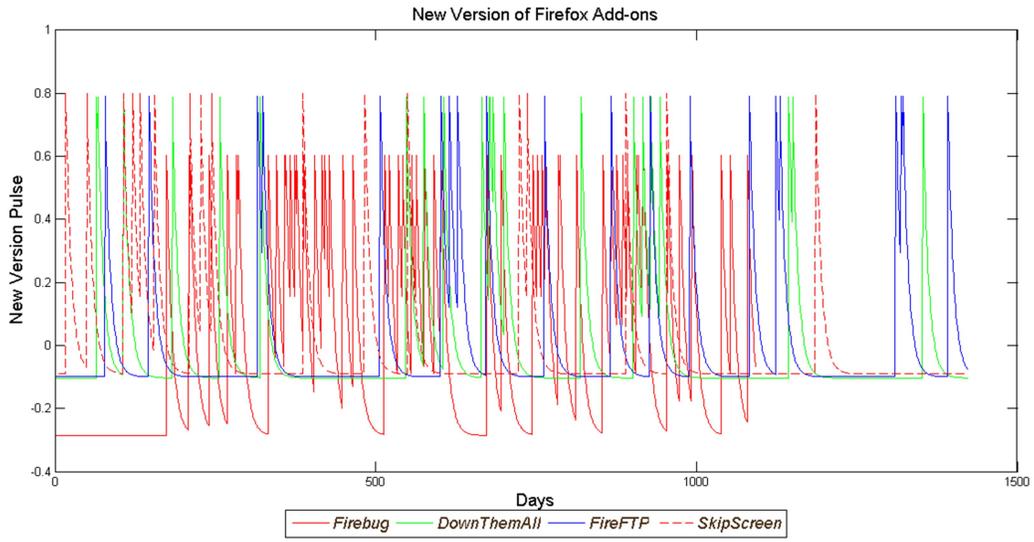

**Figure 4: One-Step Ahead Forecast of Four Sample Add-on's Cumulative Downloads**

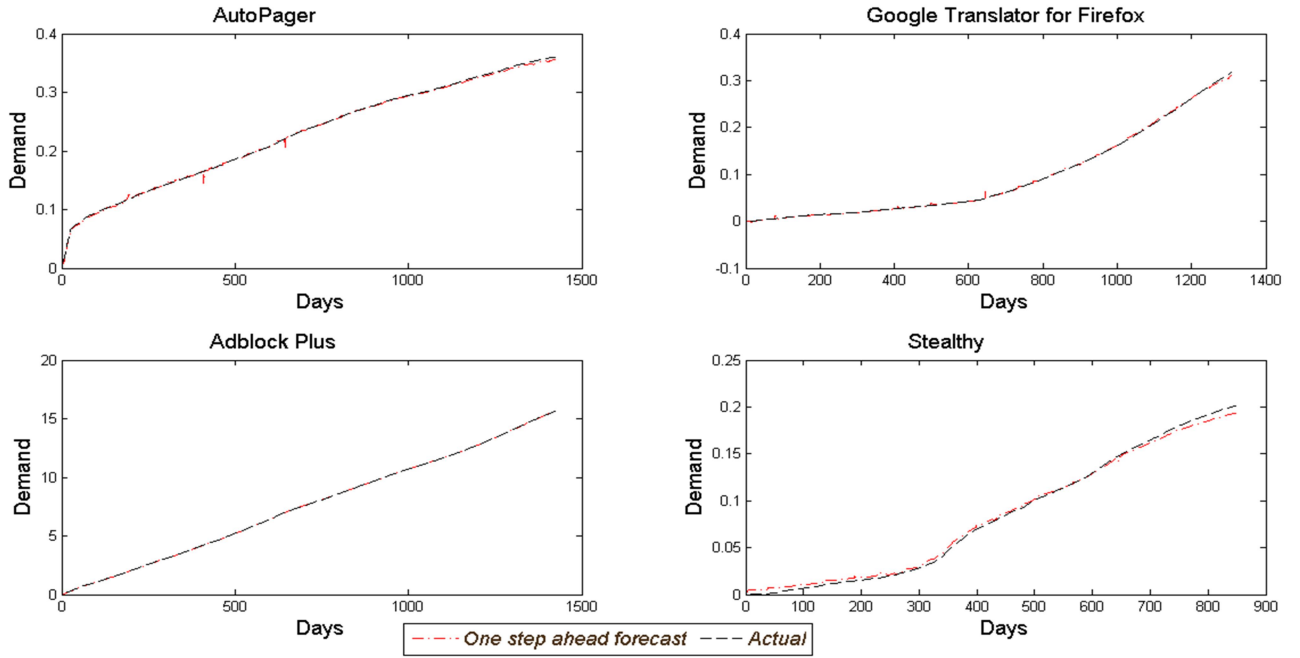

**Figure 5: One-Step Ahead Forecast of Firefox Platform Daily Users**

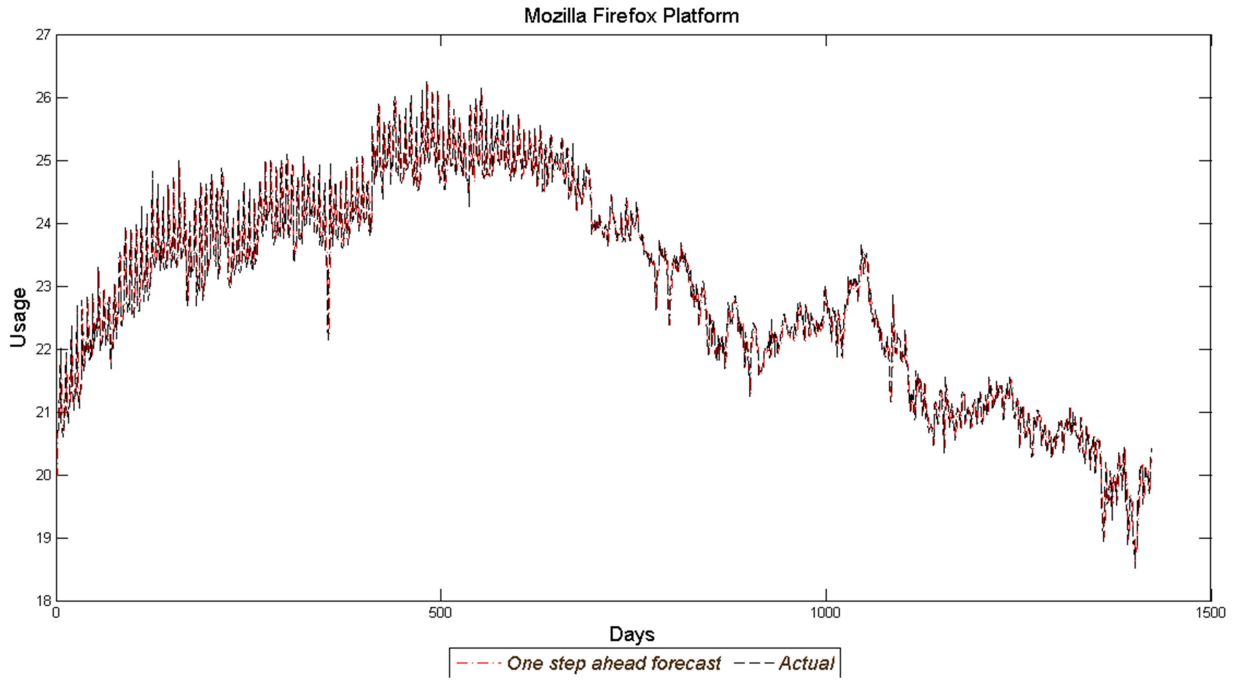

**Figure 6: Histogram of Parameter Estimate Across Add-ons**

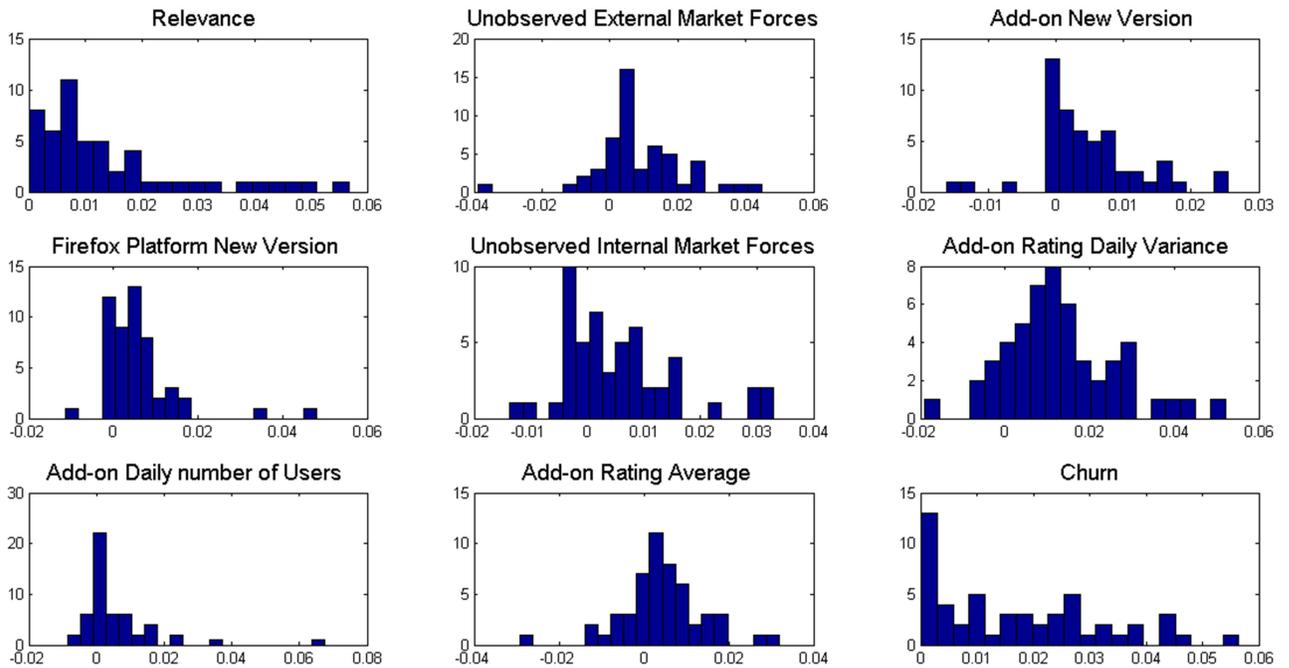

**Manuscript: The Joint Diffusion of a Digital Platform and its Complementary Goods:**

The Effects of Product Ratings and Observational Learning

**WEB APPENDIX A**

**A Test for Endogeneity**

To test for the endogeneity we extend the method for linear models in Naik and Tsai 2000 and Sonnier et al. 2012 to handle non-linear models. As discussed earlier, there is potential endogeneity in the variables: 1) Mozilla Add-on Organization (AMO) contributions, and 2) in the smoothed new release variables for Mozilla platform and its complements. We model endogeneity in the AMO contribution process in two ways. In the first, the AMO process ($Z^1$) is a latent, first order AR(1) process with a drift and a stochastic error term. The second augments this AR(1) process with instruments $Z^2$, the length of the AMO nomination queue; that is, exogenous queue length helps determine daily AMO contributions. The specification becomes:

$$\begin{aligned} y_t &= m_t + v_t \\ Z_t^1 &= \mu_t + \vartheta_t \\ m_t &= m_{t-1} + (p_0 + \mu_t p_1 + Z_t^2 p_2 + \frac{q m_{t-1}}{M_0 + A_t \kappa})(M_t - m_{t-1}) + w_t \\ \mu_t &= \gamma_1 + \gamma_2 \mu_{t-1} + \varsigma_t \end{aligned} \quad \text{(E1)}$$

where $Z_t^1$ denotes the observed AMO contribution level, and $\mu_t$ the latent AMO contribution level, and $\gamma_1, \gamma_2$ are the parameters to estimate. Again, in the second approach we augment the (AR(1)) process, the state equation of the latent AMO contribution as follows:

$$\mu_t = \gamma_1 + \gamma_2 \mu_{t-1} + \gamma_3 Z_t^2 + \varsigma_t \quad \text{(E2)}$$

We recast these models in state space form:

$$y_t = m_t + v_t \quad \text{(E3)}$$

$$\begin{pmatrix} m_t \\ Z_t^1 \end{pmatrix} = f\begin{pmatrix} m_{t-1} \\ \mu_t \end{pmatrix} + \begin{pmatrix} w_t \\ \vartheta_t \end{pmatrix} \quad \text{(E4)}$$

$$\mu_t = g(\mu_{t-1}) + \varsigma_t \quad \text{(E5)}$$

where

$$\begin{pmatrix} w_t \\ \vartheta_t \end{pmatrix} \sim MVN(0, \Sigma) \quad \text{(E6)}$$

$$v_t \sim N(0,V)$$
$$\varsigma_t \sim N(0,\Psi)$$ (E7)

Our estimation occurs in two-steps. In the first, conditional on the latent AMO contribution $\mu_t$, and the error distribution ($w_t | \vartheta_t$), we estimate $m_t$ using an Extended Kalman Filter (EKF). Then in the second, conditional on $m_t$, we use (E4) as the observation equation, (E5) as the state equation, and (E6) as the joint variance to estimate the latent AMO contribution $\mu_t$. The results of this estimation for both models are presented in the table 1A. The correlation and the off diagonal elements of the variance-covariance matrix are not significant; this suggests that endogeneity of AMO contributions are unlikely to be a problem.

We use the same approach (E1) to test for the endogeneity of the smoothed new release variables for Mozilla platform and its complements. Table 1B presents the confidence intervals of these error correlations and variance-covariance matrix elements. Similarly, the correlation and the off diagonal elements of the variance-covariance matrix suggest that there is no significant endogeneity.

**Table 1A: Potential endogeneity test for Mozilla Add-on Organizations Contributions**

| Model | Estimate | Mean | STD | 2.5% | 97.5% |
|---|---|---|---|---|---|
| Model 1 | $\text{Corr}(w_t, \vartheta_t)$ | 0.0140 | 0.1260 | -0.1980 | 0.2230 |
|  | $\Sigma_{21}$ | 0.0005 | 0.0050 | -0.0070 | 0.0080 |
| Model 2 | $\text{Corr}(w_t, \vartheta_t)$ | -0.0020 | 0.1150 | -0.1950 | 0.1930 |
|  | $\Sigma_{21}$ | -3.5e-5 | 0.0050 | -0.0080 | 0.0080 |

**Table 1B**
**Potential endogeneity test for new releases of Mozilla platform and its Add-ons**

| Add-on | Corr $(w_t, \vartheta_t^1)$ | | Corr $(w_t, \vartheta_t^2)$ | | $\Sigma_{13}$ | | $\Sigma_{23}$ | |
|---|---|---|---|---|---|---|---|---|
| | 2.5% | 97.5% | 2.5% | 97.5% | 2.5% | 97.5% | 2.5% | 97.5% |
| 1 | -0.1012 | 0.0001 | -0.1033 | 0.0139 | -0.0004 | 0.0000 | -0.0003 | 0.0000 |
| 2 | -0.0324 | 0.0320 | -0.0332 | 0.0454 | -0.0002 | 0.0001 | -0.0001 | 0.0001 |
| 3 | -0.0951 | 0.0309 | -0.0829 | 0.0623 | -0.0001 | 0.0000 | -0.0003 | 0.0001 |
| 4 | -0.0001 | 0.0000 | -0.0002 | 0.0001 | -0.0603 | 0.0216 | -0.0437 | 0.0378 |
| 5 | -0.0738* | -0.0024* | -0.0172 | 0.0592 | -0.0002 | 0.0000 | -0.0001 | 0.0001 |
| 6 | -0.112* | -0.001* | -0.051 | 0.049 | -8.e-5* | -3.e-5* | -0.0001 | 0.0001 |
| 7 | -0.038 | 0.042 | -0.037 | 0.042 | -0.0001 | 0.0001 | -0.0001 | 0.0001 |
| 8 | -0.054 | 0.025 | -0.024 | 0.063 | -0.0001 | 0.0000 | -0.0001 | 0.0001 |
| 9 | -0.023 | 0.037 | -0.046 | 0.032 | -0.0001 | 0.0001 | -0.0002 | 0.0001 |
| 10 | -0.034 | 0.035 | -0.042 | 0.042 | -0.0001 | 0.0001 | -0.0001 | 0.0001 |
| 11 | -0.063 | 0.028 | -0.024 | 0.070 | -0.0002 | 0.0000 | -0.0002 | 0.0002 |
| 12 | -0.009 | 0.090 | -0.059 | 0.034 | -0.0001 | 0.0001 | -0.0002 | 0.0001 |
| 13 | -0.060 | 0.041 | -0.041 | 0.060 | -0.0002 | 0.0001 | -0.0001 | 0.0001 |
| 14 | -0.092 | 0.006 | -0.025 | 0.067 | -0.0004 | 0.0001 | -0.0002 | 0.0002 |
| 15 | -0.041 | 0.034 | -0.015 | 0.057 | -0.0001 | 0.0000 | -0.0001 | 0.0001 |
| 16 | -0.0888 | 0.0250 | -0.0381 | 0.0690 | -0.0003 | 0.0000 | -0.0002 | 0.0002 |
| 17 | -0.0904 | 0.0387 | -0.0360 | 0.0518 | -0.0003 | 0.0001 | -0.0001 | 0.0001 |
| 18 | -0.0599 | 0.0423 | -0.0690 | 0.0474 | -0.0004 | 0.0002 | -0.0003 | 0.0002 |
| 19 | -0.0865 | 0.0402 | -0.0789 | 0.0490 | -0.0005 | 0.0002 | -0.0005 | 0.0002 |
| 20 | -0.0699 | 0.0325 | -0.0435 | 0.0520 | -0.0002 | 0.0001 | -0.0001 | 0.0001 |
| 21 | -0.061 | 0.032 | -0.040 | 0.055 | -0.0002 | 0.0001 | -0.0001 | 0.0001 |
| 22 | -0.074 | 0.024 | -0.038 | 0.061 | -0.0002 | 0.0001 | -0.0001 | 0.0001 |
| 23 | -0.273* | -0.115* | -0.084 | 0.039 | -0.0002 | -0.0001 | -0.0003 | 0.0001 |
| 24 | -0.132* | -0.010* | -0.063 | 0.049 | -0.0003 | 0.0000 | -0.0004 | 0.0002 |
| 25 | -0.097 | 0.017 | -0.066 | 0.048 | -0.0004 | 0.0001 | -0.0004 | 0.0002 |
| 26 | -0.020 | 0.054 | -0.016 | 0.057 | -0.0001 | 0.0001 | -0.0001 | 0.0001 |
| 27 | -0.094 | 0.049 | -0.059 | 0.075 | -0.0002 | 0.0000 | -0.0002 | 0.0001 |
| 28 | -0.051 | 0.028 | -0.014 | 0.075 | -0.0002 | 0.0001 | -0.0001 | 0.0001 |
| 29 | -0.081 | 0.017 | -0.037 | 0.055 | -0.0001 | 0.0000 | -0.0001 | 0.0001 |
| 30 | -0.083 | 0.008 | -0.026 | 0.074 | -0.0001 | 0.0000 | -0.0001 | 0.0001 |
| 31 | -0.060 | 0.024 | -0.030 | 0.056 | -0.0001 | 0.0000 | -0.0001 | 0.0001 |
| 32 | -0.081 | 0.017 | -0.060 | 0.023 | -0.0001 | 0.0000 | -0.0002 | 0.0000 |
| 33 | -0.010 | 0.047 | -0.068 | -0.007 | -0.0001 | 0.0001 | -0.0002 | 0.0000 |
| 34 | -0.118 | 0.082 | -0.110 | 0.099 | -0.0008 | 0.0004 | -0.0008 | 0.0004 |
| 35 | -0.043 | 0.055 | -0.058 | 0.049 | -0.0001 | 0.0000 | -0.0002 | 0.0001 |
| 36 | -0.081 | 0.045 | -0.072 | 0.057 | -0.0003 | 0.0001 | -0.0003 | 0.0001 |
| 37 | -0.047 | 0.037 | -0.034 | 0.055 | -0.0001 | 0.0001 | -0.0001 | 0.0001 |
| 38 | -0.060 | 0.022 | -0.005 | 0.077 | -0.0001 | 0.0000 | -0.0001 | 0.0001 |
| 39 | -0.043 | 0.037 | -0.023 | 0.059 | -0.0001 | 0.0000 | -0.0001 | 0.0001 |
| 40 | -0.094 | 0.052 | -0.087 | 0.060 | -0.0006 | 0.0003 | -0.0004 | 0.0002 |
| 41 | -0.069 | 0.015 | -0.017 | 0.072 | -0.0001 | 0.0000 | -0.0001 | 0.0001 |
| 42 | -0.091 | 0.014 | -0.049 | 0.061 | -0.0001 | 0.0000 | -0.0001 | 0.0001 |
| 43 | -0.063 | 0.116 | -0.098 | 0.064 | -0.0005 | 0.0005 | -0.0005 | 0.0002 |
| 44 | -0.084 | 0.026 | -0.061 | 0.028 | -0.0003 | 0.0001 | -0.0002 | 0.0001 |
| 45 | -0.058 | 0.022 | -0.035 | 0.047 | -0.0001 | 0.0000 | -0.0001 | 0.0001 |
| 46 | -0.037 | 0.044 | -0.030 | 0.064 | -0.0002 | 0.0001 | -0.0001 | 0.0001 |
| 47 | -0.102 | 0.032 | -0.073 | 0.049 | -0.0005 | 0.0001 | -0.0004 | 0.0002 |
| 48 | -0.062 | 0.037 | -0.036 | 0.060 | -0.0002 | 0.0001 | -0.0001 | 0.0001 |
| 49 | -0.045 | 0.030 | -0.012 | 0.072 | -0.0001 | 0.0000 | -0.0001 | 0.0001 |
| 50 | -0.075 | 0.026 | -0.042 | 0.069 | -0.0002 | 0.0000 | -0.0001 | 0.0001 |
| 51 | -0.1052 | 0.0038 | -0.0665 | 0.0390 | -0.0004 | 0.0000 | -0.0003 | 0.0001 |
| 52 | -0.1080 | 0.0150 | -0.0760 | 0.0453 | -0.0003 | 0.0000 | -0.0004 | 0.0002 |

[1]-Platform; [2]-add-on